\documentstyle[aps,multicol,prl,epsf]{revtex}
\newcommand{\e   }   { {\rm e} }
\newcommand{\r   }   {  {\bf r}}
\newcommand{\Cl   }   { {\rm Cl} }
\newcommand{\Na   }   { {\rm Na} }
\newcommand{\OH  }   { {\rm OH} }
\newcommand{\HH   }   { {\rm H} }

\newcommand{\B   }   { {\rm B} }
\newcommand{\m   }   { {\rm m} }
\newcommand{\dd   }   { {\rm d^3} }
\newcommand{\q   }   {  {\bf q}}
\newcommand{\I   }   { {\rm I} }

\begin{document}

\draft
\title{Polyelectrolyte Titration: Theory and Experiment}

\author{Itamar Borukhov$^{(a)}$\footnote{present address:
Department of Chemistry and Biochemistry, UCLA, 607 C. Young Dr.
East, Los Angeles, CA 90095-1569, USA. e-mail:
itamar@chem.ucla.edu}, David Andelman$^{(b)}$\footnote{e-mail:
andelman@post.tau.ac.il}, Regis Borrega$^{(a)}$, Michel
Cloitre$^{(a)}$ \\ Ludwik Leibler$^{(a)}$ and Henri
Orland$^{(c)}$\footnote{e-mail: orland@spht.saclay.cea.fr}}

\address{
$^{(a)}$ Unit\'e Mixte de Recherche CNRS/Elf Atochem (UMR 167) \\
95 rue Danton, B.P. 108, 92303 Levallois-Perret Cedex, France \\
$^{(b)}$ School of Physics and Astronomy,
Raymond and Beverly Sackler Faculty of Exact Sciences \\
Tel Aviv University, Tel Aviv 69978, Israel\\
$^{(c)}$ Service de Physique Th\'eorique, CE-Saclay,
91191 Gif-sur-Yvette, Cedex, France}

\date{May 6, 2000}

\maketitle

\begin{abstract}
Titration of methacrylic acid / ethyl-acrylate copolymers is
studied experimentally and theoretically. At low salt
concentrations, this polyacid exhibits a plateau in the titration
curve below the neutralization point. The plateau has been often
attributed to a first-order phase transition associated with
polymer conformational changes. We argue that the specific shape
of titration curves of hydrophobic polyelectrolytes is due to
electrostatics and does not necessarily require a conformation
change of the polyelectrolyte chains. We calculate the free energy
at the mean-field level and its first-order (one loop) correction
using a loop expansion.
The latter is dominated by Debye-H\"uckel--like charge-charge
correlations as well as by correlations between dissociation sites
along the polymer chain. We show that the one-loop corrections to
the free energy lead to titration curves that agree with
experiments. In particular, the model explains the decrease of the
pH at the plateau when the polymer concentration is increased or
when salt is added to the solution.
\end{abstract}
\
\
\

\section{Introduction}

In recent years the search for environment-friendly materials
has promoted the development
of numerous water-soluble polymer applications.
Polymers can be made water soluble by making
them compatible with the strong polar environment of the aqueous media, e.g.,
by introducing either charges or strong dipoles on the chains
 \cite{poly1,poly2,poly3}.

Most charged polymers (polyelectrolytes) have
a hydrophobic backbone.
This hydrophobicity induces an effective attraction between monomers
which competes with the Coulomb repulsion between charges.
As a result, hydrophobic polyelectrolytes exhibit complex behavior
including conformation changes, macro- and meso-phase separation,
self-association and aggregation \cite{complex1,complex2,complex3}.

In spite of considerable theoretical and experimental effort, many
questions remain open. In particular, the role of the coupling between
hydrophobic attractions and long range Coulomb interactions on the
physico-chemical properties of polyelectrolyte solutions is not fully understood.
In this paper, we consider the role of such coupling in solutions of
weak polyacids.

Weak acid monomers (denoted HA) can undergo dissociation
of the type
$$ {\rm HA \rightleftharpoons H^+ + A^-}$$
where A$^-$ is the charged monomer attached
to the backbone, while the dissociated
H$^+$
charge dissolves into the solution. The
dissociation/association is an equilibrium process satisfying detailed
balance.
The ionization degree determines the effective amount of charge
on the polyelectrolyte chain,
and it
depends on the pH of the solution.
At low pH the polymer is weakly charged while at
high pH a larger fraction of monomers is dissociated and the polymer charge
saturates to its maximal value.
The most visible consequence is the solubility in water:
hydrophobic polyacids can become water soluble at
high enough pH, where the polymer charge is strong
enough to overcome the chain
hydrophobicity.
In contrast to low molecular
weight acids, the charged groups of polyacids are correlated as they
are linked together along the chain. Indeed, the dissociation of
one acid group is correlated in a complex way to the position and
number of
other charged groups on the chain.
As a result, when the amount of charge on the chain
varies,
the chain conformation is affected, and in turn, influences the
dissociation of other groups \cite{complex1,complex2,khokhlov1,rafael}.
Studies of the  interplay of these competing effects
has attracted a large amount
of experimental \cite{exp1,exp2,exp3} and theoretical \cite{th1,th2,th3} interest.
Additional motivation for these efforts is related to the use of
water-soluble polyelectrolytes and in particular
alkali-swellable polymers in many industrial applications, such
as coatings, food and cosmetic industries \cite{app1,app2}.

An elementary and wide spread experimental tool to characterize
polyacids is to perform titration experiments. In these
experiments, a strong base like NaOH is added to a solution of
weakly charged polyelectrolytes. The pH of the solution and the
equilibrium dissociation depends not only the amount of added base
(like for ordinary acids \cite{titration}), but also on the
polyelectrolyte concentration and presence of salt
\cite{tit1,tit2}. The overall shape of the titration curve depends
on the nature of the polyelectrolyte which is titrated
\cite{tit1}. In the case of poly(acrylic) acid, the pH increases
steadily with the dissociation degree. In contrast, the titration
curve of poly(methacrylic) acid shows a maximum at low
dissociation. Such a non-monotonous dependence has been related to
a conformational transition. The case of hydrophobic
polyelectrolytes is still more intriguing: the titration curve
shows a large plateau where the pH is almost constant before the
neutralization point is obtained. It has been argued in the
literature that this plateau could be associated with a
first-order phase transition between collapsed and swollen states
of the polyacid chains \cite{khokhlov1,adv}.

In this paper we show that  this peculiar shape of the titration
curves can be explained without the need to rely on
conformational phase transition
of the polyacid chains.
By including the effect of correlations of charges along the chains
in the free energy,
titration curves are calculated and show a behavior similar to the curves
obtained in experiments.
In particular, we explain how the pH depends
on the polymer concentration
and the amount of added salt.
This dependence is special to polyacid solutions and is
much weaker in monomeric acid solutions.

The paper is structured in the following way: in the next section
the mean-field free energy, its one-loop correction and the
resulting titration equations and curves are presented. Then, the
experimental measurements of the titration of methacrylic acid /
ethyl-acrylate copolymers (MAA-EA) are discussed in Sec.~III, and
the comparison between theory and experiment is presented in
Sec.~IV. The full formalism relying on a field theoretical
approach will be presented in a forthcoming publication.

\section{Theory}
\label{sec:Theory}

We present first the free energy leading to
titration curves of weak polyacids  in presence of added
salt.
Our system consists of four dissociating species: water,
methacrylic acid monomers (denoted HA),
NaOH titrating
base and NaCl salt. The dissociation reactions are written as:
\begin{eqnarray}
 & {\rm H_2O \rightleftharpoons H^+ + OH^-}, \nonumber \\
 & {\rm HA \rightleftharpoons   H^+ + A^-}, \nonumber \\
 & {\rm NaOH \rightarrow Na^+ + OH^-}, \nonumber \\
 & {\rm NaCl \rightarrow Na^+ + Cl^-}. \nonumber
\end{eqnarray}
The partial dissociation of water and acid monomers is accompanied
by an energy cost of breaking the molecular bond. They are denoted as
$\Delta_1$ and $\Delta_2$, respectively (in units of the
thermal energy $k_\B T$), and they are related to
the mass action law \cite{titration} as will be detailed below.
Since NaOH is a strong base
it is fully dissociated. Similarly, the salt is fully
dissociated.

 The polymer used in the experiment is a statistical copolymer
composed randomly of two monomers:
 methacrylic acid and ethyl-acrylate. The total polymerization
 index is denoted as $N$. A
fraction $f=1/3$ of the monomers are composed of the methacrylic
acid, i.e. they are ionizable.
Since the charged monomers are distributed uniformly along the
chain, we assign a partial charge $fe$ to each monomer
(``smearing" the charges along the chain). Note that $fe$ is the
nominal charge of each monomer, while the actual charge is related
to the partial dissociation of the acid monomers.

The concentration of monomers in the solution is denoted
$c_\m$ while the concentrations of the small ions are denoted
$c_{\rm H}$, $c_\OH$, $c_{\rm Na}$, $c_{\rm Cl}$.

The base concentration added to the solution is
denoted $c_\B$ and its ratio to the MAA monomer concentration is
defined as the degree of neutralization $\gamma$

\begin{equation}
\gamma \equiv {c_\B \over  f c_\m}
\end{equation}
where $\gamma$ varies between zero (no added base) and infinity
(large base excess). It is also useful to define the degree of
ionization $\alpha$ of the acid
\begin{equation}
\alpha=\frac{\rm [A^-]}{\rm [HA]+[A^-]}=\frac{\rm [A^-]}{fc_\m}
\end{equation}
where ${\rm [A^-]}$ and  ${\rm [HA]}$ are the concentrations of
dissociated and non-dissociated monomers, respectively.
This degree of ionization is related to the pH=$-\log_{10}c_{\HH}$
of the solution via
\begin{equation}
{\rm pH=pK_A} + \log_{10} \frac{\alpha}{1-\alpha}
\end{equation}
In addition, due to charge neutrality, $\alpha$ and $\gamma$ are
related via
\begin{equation}
\alpha=\gamma + \frac{\rm [H^+]-[OH^-]}{f c_\m}
\end{equation}

The conservation of mass implies the following relation between the ion
concentrations:

$$ c_{\rm Na}= c_\B+c_{\rm Cl}$$
Note that the Cl$^-$ ions come only from the salt, while
the Na$^+$ come both from the base and the salt.  Similarly,
the OH$^-$ ions come from the dissociation of water and base, whereas
the H$^+$ ions come from the dissociation of water and monomeric acid.

\subsection{Free energy}
We denote the full free energy as $F=F_0+\Delta F$, where $F_0$ is the mean field
term and $\Delta F$ is the one-loop correction. Below we discuss
the two terms of the free energy separately.

\subsubsection{Mean-field free energy}

In a separate work, the mean-field free energy (per unit volume
and in units of $k_\B T$) is shown to be
\begin{eqnarray}\label{fMF}
  \beta F_0 = {c_\m\over N}\left(\log{c_\m w_\m}-1\right)~~
    + \sum_{j={\rm H,OH,Na,Cl}} c_j \left(\log{c_j w_j}-1\right)
    - {v\over2}c_\m^2 + {w\over6}c_\m^3 - \lambda_0 c_\m
    \nonumber \\ \nonumber \\
    - \left(c_\OH- c_\B \right)\Delta_1
    - \left(c_{\rm H} + c_\B - c_\OH\right)\Delta_2
    + \left(c_{\rm H} + c_\B - c_\OH\right)\beta e \varphi_0
    - f c_\m\log(1+\e^{\beta e \varphi_0})
\end{eqnarray}
It turns out to be identical to the usual Flory-Huggins theory of
polymer-solvent mixtures.
The factor $1/N$ in the first term  accounts for the
reduction of translational entropy of chains of $N$ monomers.
The second term represents the entropy of mixing
of the small ions.
The parameter $w_j$ is the molar volume of the
$j$ species.
The next two terms represent the short range monomer-monomer
interactions.
The hydrophobic effect is modeled by an attractive second
virial coefficient $v$ (having dimension of volume), and a repulsive third
virial coefficient $w$ (having dimension equal to volume squared).
It is introduced to avoid collapse of the chain.
The next term represents the chain conformational
entropy,
where $\lambda_{0}$ is the conformational
entropy per monomer of a free Brownian chain.

The following two terms represent the energy cost of dissociation of
water molecules and acid monomers.
We define by $\Delta_{1}$ the energy loss for each dissociation of a water
molecule

$$
{\rm H_2 O \rightleftharpoons H^+ + OH^-}
$$
and by $\Delta _{2}$ the energy loss for each dissociation of
an acid molecule
$$
{\rm HA \rightleftharpoons H^+ + A^-}
$$
The number of OH$^-$ ions coming from
dissociation of water molecules  is equal to
 the difference between the
total number of OH$^{-}$ ions and those
coming from the  NaOH base.
Similarly, the number of dissociated acid groups is equal to the
difference between the total number
of H$^{+}$ ions and the number of H$^{+}$ ions coming from the water.

Finally, the last two terms account for the electrostatic energy,
$\varphi_0$ being the electrostatic potential
in the solution.
The first one is the electrostatic energy of the
small ions, while the second
is the electrostatic free energy of partially
dissociated  monomers \cite{epjb}.

\subsubsection{One-loop correction to the free energy}

The one-loop correction to the mean-field free energy $\Delta F$
will be presented in detail in a forthcoming publication. It is
obtained by integrating over the quadratic fluctuations of the
concentration fields. The correction term to the free energy is
given (up to an additive constant) by

\begin{eqnarray}
\label{Deltaf}
\beta\Delta F &=& {1\over 2}\int {\dd \q
\over (2\pi)^{3}} \log \Sigma(q) \\
\label{Sigma}
\Sigma(q) &=& \bigl[ 1+ (wc_\m-v)c_\m ND(\eta)\bigr]
             \bigl(q^{2}+\kappa^2\bigr)
             +4\pi l_\B f^{2}A^{2} N D(\eta) c_\m \\
\label{eta}
\eta &=& {1\over 6}a^2q^2 N
\end{eqnarray}
where the Debye-H\"uckel screening length $\kappa^{-1}$ depends on
the total concentration of small ions $c_{\rm I}$ and the effective charge
of the chain $f A(1-A)c_\m $

\begin{eqnarray}
\kappa^2  &=& 4\pi l_\B  \bigl(c_{\rm I} + f A(1-A)c_\m \bigr)
    \equiv 4\pi l_\B c_{\rm eff}
\label{kappa}\\
   c_{\rm I} &=& c_{\HH} + c_\OH+c_\B+2c_{\rm Cl} \label{cI}
\end{eqnarray}
$l_\B=e^2/\varepsilon k_\B T\simeq 7$\,A is the Bjerrum length.
The dissociation fraction of the acid monomers is given by

\begin{eqnarray}\label{A}
A &=& {\e^{\beta e \varphi_0}\over 1+\e^{\beta e \varphi_0}}
\end{eqnarray}
The Debye function $D(\eta)$ \cite{poly2} entering eq.~(\ref{Sigma}) is given by
\begin{equation}
\label{D_eta}
D(\eta)=\frac{2}{\eta}(1+\frac{\e^{-\eta}-1}{\eta})
\end{equation}

 In addition to
the small ion contribution to the electrostatic screening,
eq.~(\ref{kappa}) includes a term proportional to $A(1-A)$, where
$A$ is defined in eq.~(\ref{A}) above. This term accounts for
changes in the dissociation degree of monomers depending on the
{\em local} electrostatic potential. At small $A$ values, $A(1-A)
\approx A$, and the polymer contribution is the same as that of
disconnected monomers. However, as $A\to 1$  there is a
substantial reduction (by the factor $1-A$)
of the effective
polymer charge that contributes to screening
\cite{rafael,epjb}. This reduction can be understood in terms of
charge correlation along the chain.
Dissociated monomers prevent further dissociation
of other monomers.

In our model, the concentrations $c_{\Na}$, $c_{\Cl}$
and $c_\m$ are fixed by the amount of NaCl salt, NaOH base and
polymer in the solution while the concentrations of
dissociated H$^+$ and OH$^-$ ions depend on the degree of dissociation
of water molecules
and acid monomers (through the electrostatic potential $\varphi_0$).
Thus $c_{\HH}$, $c_\OH$ and $\varphi_0$ are variational parameters, determined
by the requirement that the free energy $F=F_0+\Delta F$ is an extremum.
In titration experiments~~$c_{\HH}$ can be monitored directly through the pH
of the solution.

\subsubsection{Simplified free energy}

In order to better understand the main contributions
to the correction $\Delta F$ let us return to the expression for $\Sigma(q)$,
eq.~(\ref{Sigma}).
First, we note that the Debye function is bound between 0 and 1 (see also
eq.~(\ref{simple1})).
For typical  values of the physical parameters (see also Sec. IV),
\begin{eqnarray}
v\simeq 25{\mbox A}^3 ~~~&~~~~ w \simeq 100 {\mbox A}^6 \nonumber\\
N\simeq 10^4 ~~~&~~~~ c_\m \simeq 1 {\mbox  {mM}}
\end{eqnarray}
it is easily seen that
for long chains ($N\gg1$), the Debye function can be approximated by
$ND(\eta)\simeq 12/a^2q^2$, and

\begin{equation}\label{solvent}
    |wc_\m-v|c_\m ND(\eta) \ll 1
\label{small1}
\end{equation}
Thus, the effect of the solvent is small
and will be neglected in this section

\begin{equation}
  \Delta F \simeq {1\over 2}\int{\dd \q\over(2\pi)^{3}}
  \log\left[q^2+\kappa^2 + z \right]
\end{equation}
where
\begin{equation}
   z= 4\pi l_\B f^{2}A^{2} N D(\eta) c_\m
\label{small2}
\end{equation}
For the same range of parameters $\kappa^2\gg z$ and $\Delta F$ can be
expanded to first order in $z$, $\Delta F = \Delta F_0 + \Delta F_1$:

\begin{equation}
  \Delta F_{0} = {1\over 2}\int{\dd \q\over(2\pi)^{3}}
        \log\left[q^2+\kappa^2\right]
\end{equation}
The integral can be calculated by including a cut-off at large $q$,
which cancels out the Coulomb self-energy, yielding

\begin{equation}\label{DeltaFel}
  \Delta F_{0} = -{\kappa^3\over 12\pi}
\end{equation}

This result is the well
known Debye-H\"uckel correlation energy of an electrolyte, because
the polymer contribution  is negligible and does not appear
in this leading term.
The correction $\Delta F_1$ is given by

\begin{equation}\label{DeltaFpol}
  \Delta F_1 \simeq {1\over 2}\int{\dd \q\over(2\pi)^{3}}
   {z \over q^2+\kappa^2}~~~ \simeq ~~~ 6f^2 A^2{l_\B\kappa^{-1}\over a^2}c_\m
\end{equation}

This correction is due to correlations along the chain between
dissociated monomers as can be seen from the following simple
argument.
Consider the screened electrostatic interaction between charged monomers
on a single infinite chain.
For a specific chain configuration the Coulomb energy per monomer is

\begin{equation}\label{Uel}
  U_{\rm el} = \sum_{j\neq 0} {f^2A^2e^2\over\varepsilon|\r_j|}\e^{-\kappa r_j}
\end{equation}
where $r_j$ is the spatial distance between the $j=0$ monomer
and another $j=\pm 1,\pm 2,...$ monomer.
For a Gaussian random walk, the typical (most probable) distance between monomers
is given by
\begin{equation}
  \langle r_j \rangle = \left({2j\over3}\right)^{1/2}a
\end{equation}
Assuming that $r_j$ can be approximated by
$\langle r_j \rangle$  in eq.~(\ref{Uel}), and
$\kappa a\ll 1$, the sum over $j$ can
be replaced by a continuous integral
leading to eq.~(\ref{DeltaFpol}).

\subsection{Titration equations}

The free energy depends on the species concentrations:
$c_\OH, c_\Na, c_\HH, c_\m, c_\Cl$. However, the concentration
of monomers, Na$^+$ and Cl$^-$, is fixed by the amount of polymer, base and salt
added to the solution. On the other hand, the concentration of H$^+$ and OH$^-$
ions is determined self-consistently by minimizing the full one-loop
free energy
$F=F_0+\Delta F$, eqs.~(\ref{fMF}), (\ref{Deltaf}),
with respect to $c_\HH, c_\OH$ and $\varphi_0$.
The resulting equations of state determine the dependence of the pH of the solution on
the other system parameters.

\begin{eqnarray}
&&\log ({c_\HH \omega_\HH}) + \beta e \varphi_0 -\beta \Delta_2 + 2 \pi l_\B
I_1 = 0 \label{H1}\\
&&\log ({c_\OH \omega_\OH}) - \beta e \varphi_0 -\beta \Delta_1 + \beta
\Delta_2  + 2 \pi l_\B I_1 = 0 \label{OH1}\\
&&c_\HH+c_\B -c_\OH -f c_\m A \biggl( 1 - 2 \pi l_\B (1-A)
\bigl((1-2A) I_1 +2 A N f I_2 \bigr)\biggr)=0 \label{phi01}
\end{eqnarray}
where the quantities $I_1$ and $I_2$ are defined as:
\begin{eqnarray}
\label{I11}
I_1&=& \int_0^{\Lambda} \frac{k^2 dk}{2 \pi^2}
\ \frac{1+Nc_\m(wc_\m-v)D(\eta)}{\Sigma(k)} \\
\label{I21}
I_2&=& \int_0^{\Lambda} \frac{k^2 dk}{2 \pi^2}
\ \frac{D(\eta)}{\Sigma(k)}
\end{eqnarray}
and $\Lambda$ is a short distance (large $q$) cut-off.

With the full expression (\ref{D_eta}) of the Debye function $D(\eta)$,
the integrals in
(\ref{I11}), (\ref{I21}) cannot be performed analytically. However,
for the values of the parameters
used in the experiments, namely polymerization index $N\sim 10^{4}-10^{5}$
and monomer length $a\sim 5$A, it is possible to use a simplified form
for the Debye function:
\begin{equation}
\label{simple1}
D(\eta) \simeq \frac{1}{1+\eta/2}
\end{equation}
 As can easily be seen by comparing eq.~(\ref{simple1})
with the exact expression (\ref{D_eta}), the
above form has the right behavior both at small and large values of $\eta$.
Within this approximation, the integrals are given by:
\begin{eqnarray}
\label{I1f1}
I_1&=& -\frac{1}{2\pi R^2}\frac  {R^2 \kappa^2
\kappa_+ \kappa_-/2 + (\lambda+1) \kappa^2 +u}{\kappa_+ \kappa_-
(\kappa_+ + \kappa_-)} \\
I_2&=&\frac{1}{2\pi R^2 (\kappa_+ + \kappa_-)}
\end{eqnarray}
where we have used the notation:
\begin{eqnarray}
\lambda &=& N c_\m (wc_\m-v) \\
u &=& 4 \pi l_\B A^2 N f^2 c_\m \\
R^2 &=& N a^2/6
\end{eqnarray}
and
\begin{equation}
\kappa_{\pm}^2= \frac{1}{R^2} \biggl(\lambda +1+\frac{R^2}{2}\kappa^2 \pm
\sqrt{\bigl(\lambda+1-R^2 \kappa^2/2\bigr)^2 - 2uR^2} \biggr)
\end{equation}
Note that in eq.~(\ref{I1f1}) we have omitted a term, equal to $\Lambda /
2\pi^2$, which exactly cancels
the Coulomb self-energy.

With these definitions, eqs.~(\ref{H1}) and (\ref{OH1})
can be recast in the form:

\begin{equation}
c_\HH \cdot c_\OH = 10^{-14} \,\e^{-4 \pi l_\B I_1} \label{water1}
\end{equation}
which shows the change induced by fluctuations on the water dissociation constant
and
\begin{equation}
c_\HH = \frac{1-A}{A} \ 10^{\rm -pK_A}\, \e^{-2 \pi l_\B I_1} \label{CH1}
\end{equation}

These two equations, together with eq.~(\ref{phi01}),
which expresses charge
neutrality at the one-loop level, are solved iteratively. The numerical solution
is obtained by using the mean field values as a starting point for the iterations,
and convergence is usually achieved after a few iterations.

\subsection{Structure Function --- $S(q)$}

In scattering experiments the structure function $S(q)$ is readily
obtained. It is related to the Fourier transform of the
  various density-density correlations in the sample. For example,
  we can regard the monomer-monomer correlations:
\begin{equation}
   S(q) =\langle \delta c_\m(q) \delta c_\m(q) \rangle / c_\m
\end{equation}

Since the one-loop expansion takes into account Gaussian
fluctuations, it can also be used to calculate the structure
function. The result is
\begin{equation}
   S^{-1}(q) = {1\over N D(\eta)} + wc_\m^2 - vc_\m
    + {f^2 A^2 4\pi l_\B c_\m \over q^2 + \kappa^2}
\end{equation}
The inverse structure function $S^{-1}(q)$ is the energy penalty
associated with density fluctuations at a wavenumber $q$. A
minimum in $S^{-1}(q)$ corresponds to the strongest fluctuating
wavenumber $q^*$. As a result, incoming radiation at this
wavenumber interacts most strongly with the sample and a peak
appears in the structure function and consequently in the
scattering intensity.

An instability appears when $S^{-1}(q)$ vanishes, corresponding to
a divergence of the peak in the structure function. When the
instability appears at a finite $q=q^*$ the system undergoes a
meso-phase separation and becomes spatially modulated. If,
however, the instability is at $q=0$ the system undergoes a
macrophase separation. In our case this occurs when
\begin{equation}
    vc_\m = wc_\m^2 + {1\over N} + {f^2 A^2 c_\m \over c_{\rm eff}}
\end{equation}
Recall that $c_{\rm eff}$ and $A$ depend on the pH and the degree of
neutralization.

\section{Experiment}
\label{sec:Experiment}

\subsection{Preparation of samples}
The copolymer polyelectrolyte chains used in this study are
prepared by standard emulsion
polymerization techniques using neutral ethyl-acrylate
(EA) monomers and methacrylic
acid (MAA) monomers. The methacrylic acid is a weak acid with pK$_{\rm A}$=4.5.
The weight fraction of MAA in the   copolymer
is equal to 0.35 taken to be $f=1/3$ in the theoretical section.
The emulsion polymerization is
performed under starved monomer conditions. Under these conditions,
the MAA and EA monomers
are evenly distributed along the polymer chain. From size exclusion chromatography,
we have estimated the molecular weight of the chains to be of the
order of 106 daltons, corresponding to a polymerization index
$N\simeq 10^4$. The MAA-EA copolymers
precipitate
from the aqueous solution,
as they are insoluble in water. The precipitate is carefully
washed by ultrafiltration in order to
remove surfactants, unreacted monomers and initiators.
This cleaning procedure is stopped
when the resistivity of the water flushed through the separation membrane is that of
pure water (18.2 M$\Omega$/m).
The solid content of the stock solution is then determined accurately by
drying and weighing.

\subsection{Titration measurements}
Polymer solutions at different weight concentrations are prepared by mixing weighted
amounts of the stock solutions with de-ionized water.
The dissolution of CO$_2$ is prevented
by carefully de-gassing the solution with nitrogen.
Each of the polyelectrolyte solutions is
neutralized by an NaOH solution with a molar concentration
ranging from 0.1M to 2M
depending on the polymer concentration. Upon neutralization, the methacrylic acid is
neutralized and repulsive forces due to the negative charges cause the chain to expand,
resulting in the progressive solubilization of the polymer chains.
As a result the solution
becomes transparent and its viscosity increases.
In the following, we shall characterize
the neutralization of the polymer chains by the degree of
neutralization, $\gamma=c_\B/(fc_\m)$, which is the
ratio of the amount of added base to the amount of available acids groups.
Titration experiments are performed
using a pH-meter (Metrohm 691) with a combined
glass electrode. The measurements are made at 20$^o$C with constant
stirring under
a nitrogen atmosphere.
In parallel, the conductivity of the solution is measured
(Metrohm 712).

The measurements reveal that all small ions present in solution
are free and contribute to the conductivity. This was checked at
different $\gamma$ values by changing the polymer concentration.
The measured conductivity is the exact sum of the conductivities
coming from the  Na$^+$  ions (whose concentration is known from
the amount of added NaOH base), and H$^+$, OH$^-$ (known from the
pH), while the small contribution of the polymer is negligible.
These conductivity measurements demonstrate that the distribution
of small ions is homogeneous in the solution, and no evidence for
the Donnan effect and counter-ion condensation is observed.

\section{Results: comparison of experiments with theory}

In this section we present the experimental results for the titration
curves of MAA-EA copolymers
and compare them with the theory of Sec.~II.
Fig.~1(a) shows the titration curves measured for different polymer
concentrations. They differ substantially from the titration curves of monomeric
methacrylic acid (MAA). Methacrylic acid in aqueous solution has the
typical behavior of a weak
monomeric acid. The pH increases monotonously with $\gamma$, takes
the value pK$_{\rm A}=4.5$ for
$\gamma$=0.5 and then jumps near $\gamma$=1. By contrast, the titration
curves of MAA-EA copolymers
exhibit the following behavior: in the range
$0\lesssim \gamma \lesssim 0.2$, the pH
increases sharply to a plateau
value which remains nearly constant up to $\gamma\approx 1$ where the
amount of added base equals that of the monomer (neutralization point).
The jump of the pH  for $\gamma\lesssim 0.2$ is associated with a
swelling phase transition of the latex particles
i.e. a change of conformation of the chains. A similar phenomenon
has been observed during the
titration of pure poly(methacrylic acid) \cite{scat2}.

Let us mention that it is possible to calculate the swelling
transition of the polymer as function of monomer concentration,
$c\m$. This can be done by minimizing the free energy with respect
to  $c_\m$ (in addition to the other annealed degrees of freedom
discussed above). Any non-convexity of the free energy signals the
existence of a polymer precipitate in excess water.

On Fig.~1 several titration curves are plotted
for different concentrations
of the same
MAA-EA copolymer. As the concentration $c_\m$ increases,
the jump in the pH at
$\gamma\approx 1$ increases. Note that
the deviation between the pH
of the MAA monomeric acid
and the MAA-EA copolymers ($c_\m=0.1$~M
in Fig. 1),
is large in the plateau region ($\gamma <1$).
On the other hand, the deviation
is quite small for $\gamma>1$ when the acid is almost completely dissociated.
The value of the pH at the
plateau depends strongly on the polymer concentration: the larger the $c_m$,
the lower the plateau.
It is surprising that even though MAA-EA copolymers contain carboxylic groups,
the plateau value of the pH below $\gamma=1$ may be neutral or even greater than 7.

This difference in behavior between the MAA monomers and MAA-EA copolymers
might be associated with a
complex structure of the MAA-EA copolymer
for $\gamma \lesssim  1$.
Collapsed microdomains may exist on the chains due to the competition between
the hydrophobic attraction and Coulomb repulsion. The plateau region
suggests the existence of such collapsed microdomains along the chains. For
$\gamma \gtrsim 1$ the chains are in a swollen state and their behavior resembles
that of monomeric weak acid.

In Fig. 1(b) we plot for comparison the titration curves as
calculated from the theory. The titration equations presented in
Sec. II.B are solved numerically by an iteration procedure
starting from the mean field values as the first iteration and
including the corrections of the second iteration.
Since there are a
couple of unknown physical parameters (like the second and third
virial coefficients, $v, w$ and the monomer size $a$), we do not
try to fit the experimental titration curve. Rather, we note that
the corresponding theoretical curves look very similar to the
experimental ones, demonstrating the same type of plateau for low
$\gamma$ and the same trend with the monomer concentration.


Figure 2 shows  the effect of adding a monovalent salt (NaCl).
The overall shape of the
titration curves remain unchanged. However, the value of the pH at the
plateau increases upon the addition of salt, while the pH for $\gamma \gtrsim 1$
is essentially independent on the salt concentration.
This indicates that neutralization
of the carboxylic groups carried by the chains becomes easier as the ionic
strength is increased.  As the Coulomb interaction is screened by the salt,
there is less
correlation between charged groups along the chain
when their distance is larger than the Debye length. Therefore, monomers separated by
distances larger than the Debye length can dissociate independently from each other.
For $\gamma \gtrsim 1$, almost all the charged groups on the chains are dissociated;
thus,
the salt has no effect on the pH. The comparison between experiment and theory is
presented in part (a) and (b), where a good agreement can be observed. Note
the larger deviation from the experimental results
for higher salt concentration ($c_{\rm NaCl}=0.1\,$M).

In Fig.~3, the plateau value of the pH is plotted on a semi-log plot
as function of the ionic strength
$c_\I=c_\Na+c_\HH+c_\Cl+c_\OH$.
The plateau value of the pH is taken at $\gamma=0.5$.
It is interesting to note that the different points
taken at various salt and base concentrations collapse on a single curve.
This result indicates that the pH is dominated by electrostatic
effects and not by conformation changes of the chains.
When the ionic strength increases, the fixed charges carried by the polymer are
screened and the energy associated with the dissociation of
a carboxylic group decreases.

This behavior can be characterized semi-quantitatively by noting
that when $\gamma\simeq 1/2$ the pH is much higher than the
pK$_{\rm A}$ and, therefore,
 $\gamma^{(0)}\to 1$ and $A\to 1$.
Consequently, at the plateau $\Delta\gamma\simeq 1/2$. To first
order in $10^{\rm pK_A-pH}$ the plateau value can be obtained from
the simplified free energy, Sec. II.A.3 and reads:
\begin{eqnarray}\label{pH_plateau}
  {\rm pH}_{\rm plateau} \simeq {\rm pK_A} +
    \log_{10}\left[ \kappa l_\B
                    + 24\pi f^2c_m {l_\B^2\kappa^{-3}\over a^2}
            + 24 f {l_\B\kappa^{-1}\over a^2} \right]
\end{eqnarray}
Recall that $\kappa$ depends on the total amount of ions in the
system through eq.~(\ref{kappa}). The above expression is used for
the plot of the solid curve of Fig.~3b, where the pH of the
plateau is shown as a function of the effective ionic strength
$c_{\rm eff}$ defined in eq.~\ref{kappa}. In this regime, the
difference between $c_{\rm eff}$ and  $c_{\rm I}$ is a small
correction of order $10^{\rm pK_A-pH}$.

Equation~\ref{pH_plateau} exhibits three different regimes
depending on the relative importance of the various contributions
to the logarithmic term. In each regime, $c_\HH$ has a different
power law dependence on the total ionic strength
 $c_{\rm eff}=c_{\rm I} + fA(1-A)c_m$ leading to different slopes in Fig.~3.
At low ionic strength the second term dominates
and $c_\HH\propto c_{\rm eff}^{-3/2}$ (long dashed line in Fig.~3b).
At intermediate ionic strength the last term dominates and
 $c_\HH\propto c_{\rm eff}^{-1/2}$ (short dashed line).
Finally, at high ionic strength (beyond the experimentally accessible values)
 $c_\HH\propto c_{\rm eff}^{1/2}$.

In Figure~4 the effect of polymer concentration and ionic strength
on $S(q)$ is shown. The structure function is calculated at the
plateau regime were $A\simeq 1$. The dependence on the controlled
system parameters: the pH, the salt concentration and the degree
of neutralization $\gamma$ is taken implicitly into account in
$c_{\rm eff}$. At low values of $c_{\rm eff}$, the structure
function exhibits a peak at finite $q$ corresponding to the most
favorable density--density fluctuations. As $c_{\rm eff}$
increases, $S(q\to 0)$ increases until the peak disappears. This
increase is accompanied by an increase in the osmotic
compressibility of the solution.

The effect of varying the polymer concentration $c_\m$ at a fixed
$c_{\rm eff}$ is depicted in the inset. The peak is stronger at
lower polymer concentrations indicating that fluctuations become
considerably stronger at lower concentrations. This effect agrees
well with the titration curves shown in Fig.~1. Indeed, at low
polymer concentrations the shift in the titration curve is
stronger than at high concentrations. The  peak position shifts to
higher wavenumbers (smaller length scale) at higher concentration,
similar to the correlation length of polymer solutions that
decreases with increasing polymer concentration.

\section{Conclusions}
\label{sec:Conclusions}

We have presented an experimental study of the titration of weak
  polyacids by a strong base.
Our experimental findings performed
on solutions of MAA-EA copolymers,
are supported by theoretical calculations,
which show similar trends of the pH variation with the
concentration of added base, salt and
polyacid.

Titration experiments are one of the simplest and most
useful experimental
tools to probe the degree of neutralization of monomeric and polymeric
acids. Since titration curves of weak acids are universal, any
deviation from this behavior, as observed here, can be the
signature of a
complex behavior coupling Coulombic and hydrophobic interactions.

The most striking feature is the existence of a plateau of
the pH for low degree of neutralization. The plateau in the pH is at higher value
than the corresponding pH of the monomeric acid indicating that the charges on the chain
inhibit the dissociation of other charged groups. The pH is not affected strongly by
further addition of the base till the neutralization point $\gamma=1$.
This is probably due to the existence of collapsed microdomains along the chains for
$\gamma \lesssim 1$. The pH at the plateau decreases as function of the ionic strength.
This demonstrates that non-specific electrostatic interactions are responsible for
the existence of the plateau, since the pH depends mostly
on the amount of small ions and not on their type
(for a fixed degree of neutralization).

The theory presented above includes one-loop corrections to the mean-field free energy.
On the mean-field level, there is no difference between the polymeric and monomeric
titration curves except for the  translational entropy of the chains.
The one-loop correction couples the chain connectivity
with the electrostatics and induces the large deviations in the titration curves.
We modeled the polyacid as flexible chains using the standard Debye function.
Our formalism can also be applied to different models of chain
elasticity. In particular, for the case of semi-flexible chains
the same titration equations are obtained with a modified Debye
function taking into account the persistence length of the chains.

Finally, it will be interesting to complement this study by scattering
experiments where
it might be possible to resolve the chain microstructures and relate them
to the degree of ionization of the chains.

\bigskip

\noindent
{\em Acknowledgments:}

IB gratefully acknowledges the support of the
Chateaubriand postdoctoral fellowship and the hospitality
of the Elf-Atochem research center at Levallois-Perret.
DA acknowledges partial support from the US-Israel
Binational Foundation (BSF)  under grant No. 98-00429, and the
Israel Science Foundation founded by the Israel Academy of
Sciences and Humanities --- centers of Excellence Program.



\pagebreak

\begin{figure}
{\bf Fig.~1}~~~{pH plotted as a function of the degree of
neutralization $\gamma$ with no added salt. In (a) the
experimental titration curves are shown for different polymer
concentration: $c_\m=1$\,mM (10$^{-4}$ g/g) full circles;
$c_\m=10$\,mM (10$^{-3}$ g/g) full triangles; $c_\m=0.1$\,M
(10$^{-2}$ g/g) full squares. The open circles denote the
titration curve of the monomeric MAA at $c_\m=0.1$\,M (10$^{-2}$
g/g). In (b) the theoretical results are plotted for the same
polymer concentrations: $c_\m=1$\,mM (dotted curve); $c_\m=10$\,mM
(short dashes); $c_\m=0.1$\,M (long dashes). The solid line
corresponds to the mean field solution (no correlations and no
connectivity) for $c_\m=0.1$\,M. It reproduces nicely the
conventional monomeric titration curve. The polymer parameters we
take are: $a=5$\,A, $f=1/3$ and pK$_{\rm A}=4.5$.
}
\end{figure}

\begin{figure}
{\bf Fig.~2}~~~{pH plotted as a function of the degree of
neutralization $\gamma$ for different amount of added salt. The
experimental values in (a) are compared with the theoretical ones
in (b). The polymer concentration is $c_\m=1$\,mM (10$^{-4}$ g/g).
The concentration of NaCl in the solution is: $c_{\rm NaCl}=0$
(open circles/solid line), 1\,mM (full circles/dotted curve),
10\,mM (full diamonds/short dashes), 0.1\,M (full squares/long
dashes). The theoretical curves are calculated using the same
parameters as in Fig.~1.}
\end{figure}

\begin{figure}
{\bf Fig.~3}~~~{Semi-log plot of the pH at the plateau as a
function of the ionic strength. Comparison between experimental
results (a) and theoretical calculation (b). The experimental
measurements are plotted in (a) as function of the total amount of
small ions $c_{\rm I}$. The open circles give the pH of salt-free
solutions at different monomer concentration extracted from Fig.
1. The full symbols give the pH of solutions with increasing
amount of added salt: $c_\m=1$\,mM (10$^{-4}$ g/g) full circles,
and $c_\m=0.1$\,M (10$^{-2}$ g/g) full triangles. The data
collapse indicates that the pH depends only on the total ionic
strength. The theoretical results are plotted in (b) as a function
of $c_{\rm eff}=c_{\rm I}+fA(1-A)c_\m$ using the same parameters
as in Fig.~1. See text for details of the approximations made in
the calculation. }
\end{figure}

\begin{figure}
{\bf Fig.~4}~~~The structure function $S(q)$ as function of the
wavenumber $q$. The structure function is calculated at the
plateau of the titration curve where $A\simeq 1$. The physical
parameters used in the calculation are
 $a=5$\,A, $v=25$\,A$^3$, $w=100$\,A$^6$, $N=10^4$, $f=1/3$ and pK$_{\rm A}=4.5$.
The monomer concentration is $c_\m=$\,1mM while the effective ionic strength
is $c_{\rm eff}=0.01$\,mM (solid line), $c_{\rm eff}=1$\,mM (dotted curve) and
 $c_{\rm eff}=0.1$\,M (dashed curve).
In the inset the effective ionic strength is fixed at $c_{\rm
eff}=1$\,mM while the polymer concentration is varied:
 $c_\m=1$\,mM (solid line),
 $c_\m=10$\,mM (dotted curve)
and $c_\m=0.1$\,M (dashed curve).
\end{figure}

\pagebreak

\begin{figure}[tbh]
  \epsfxsize=0.4\linewidth
  \centerline{\hbox{ \epsffile{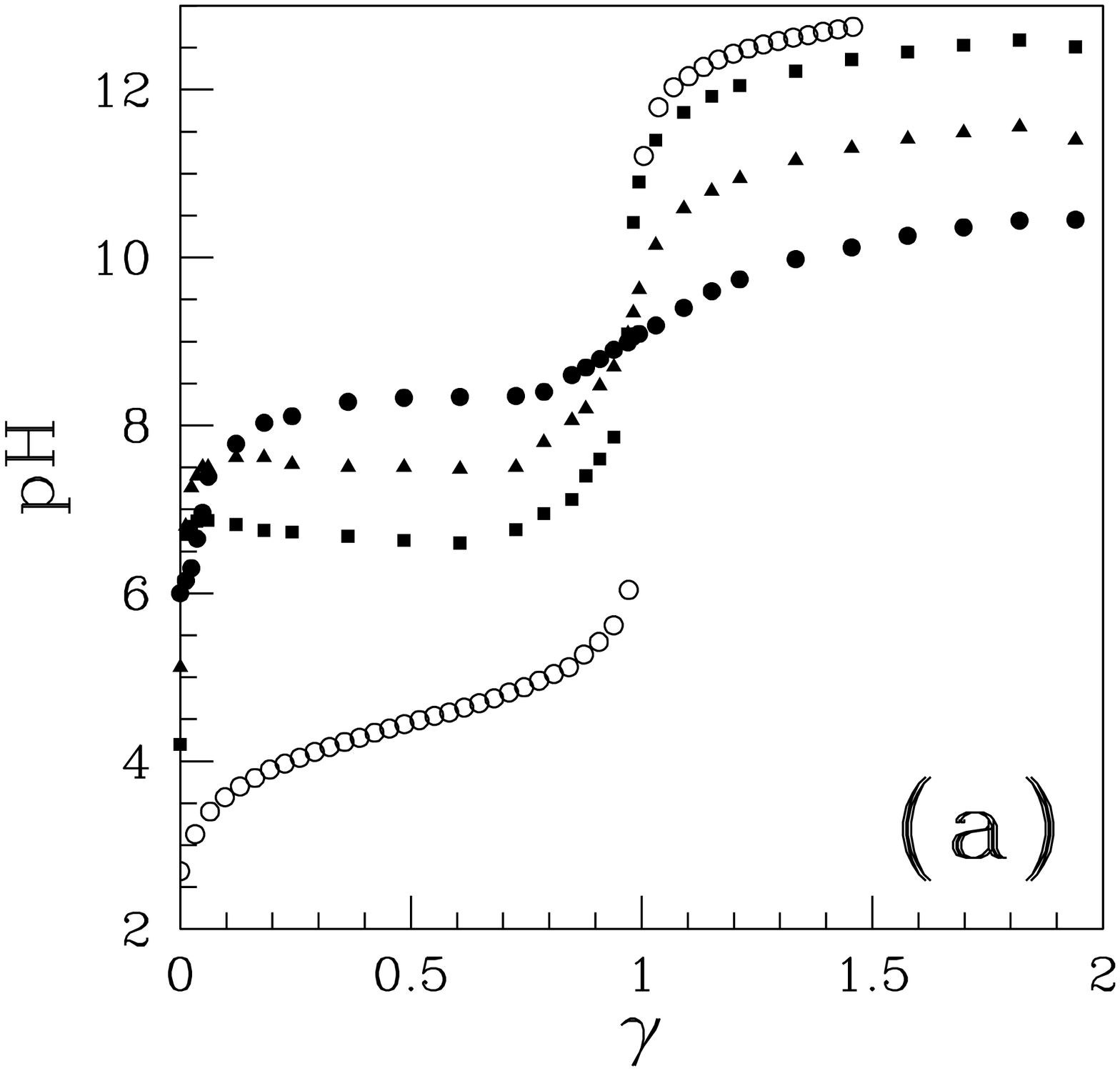}}
  \hfill
  \epsfxsize=0.4\linewidth
              \hbox{ \epsffile{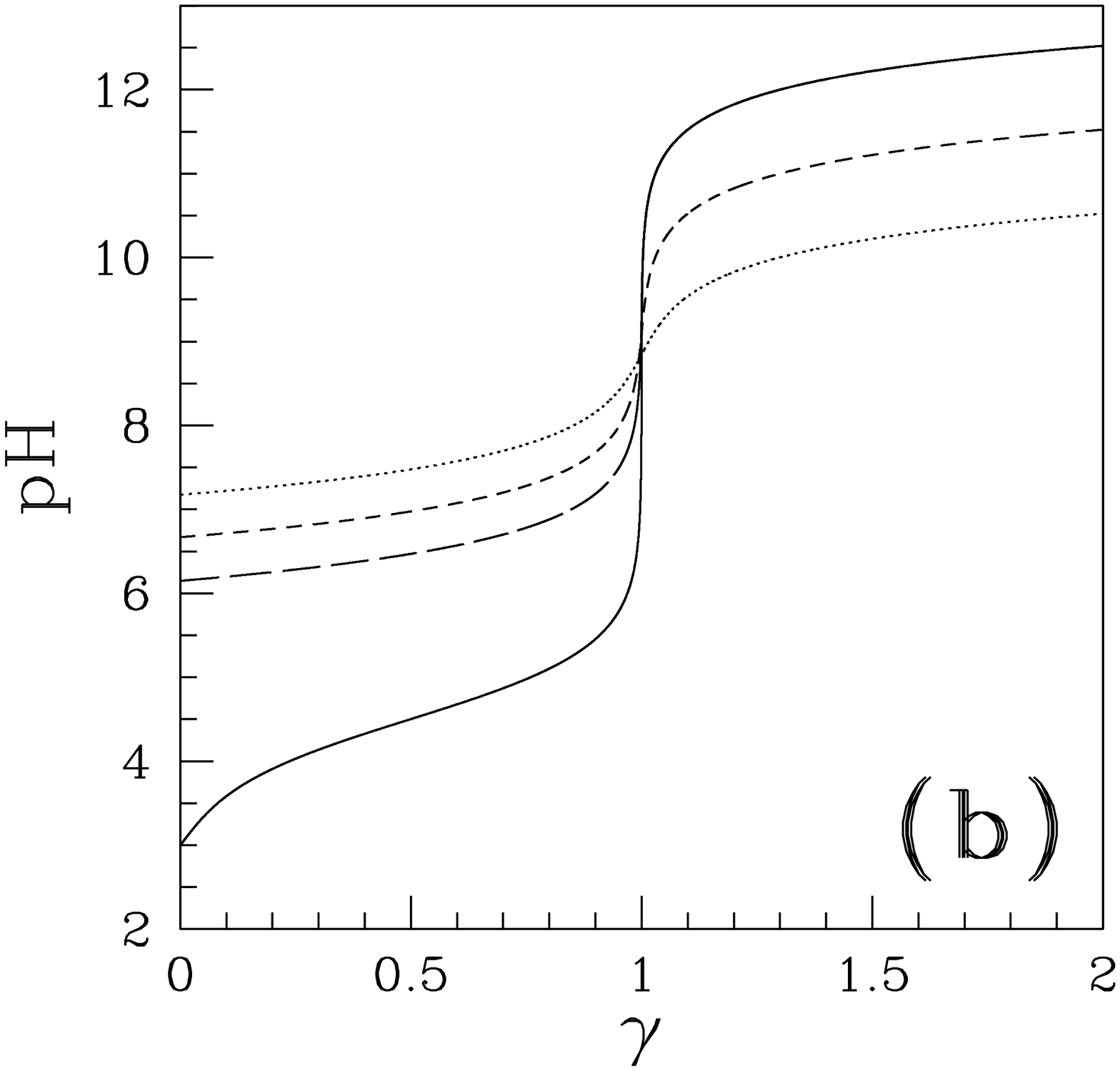}}}
\end{figure}
\vspace{1cm}
{\large Borukhov et al: Fig.~1}
\vfill

\begin{figure}[tbh]
  \epsfxsize=0.4\linewidth
  \centerline{\hbox{ \epsffile{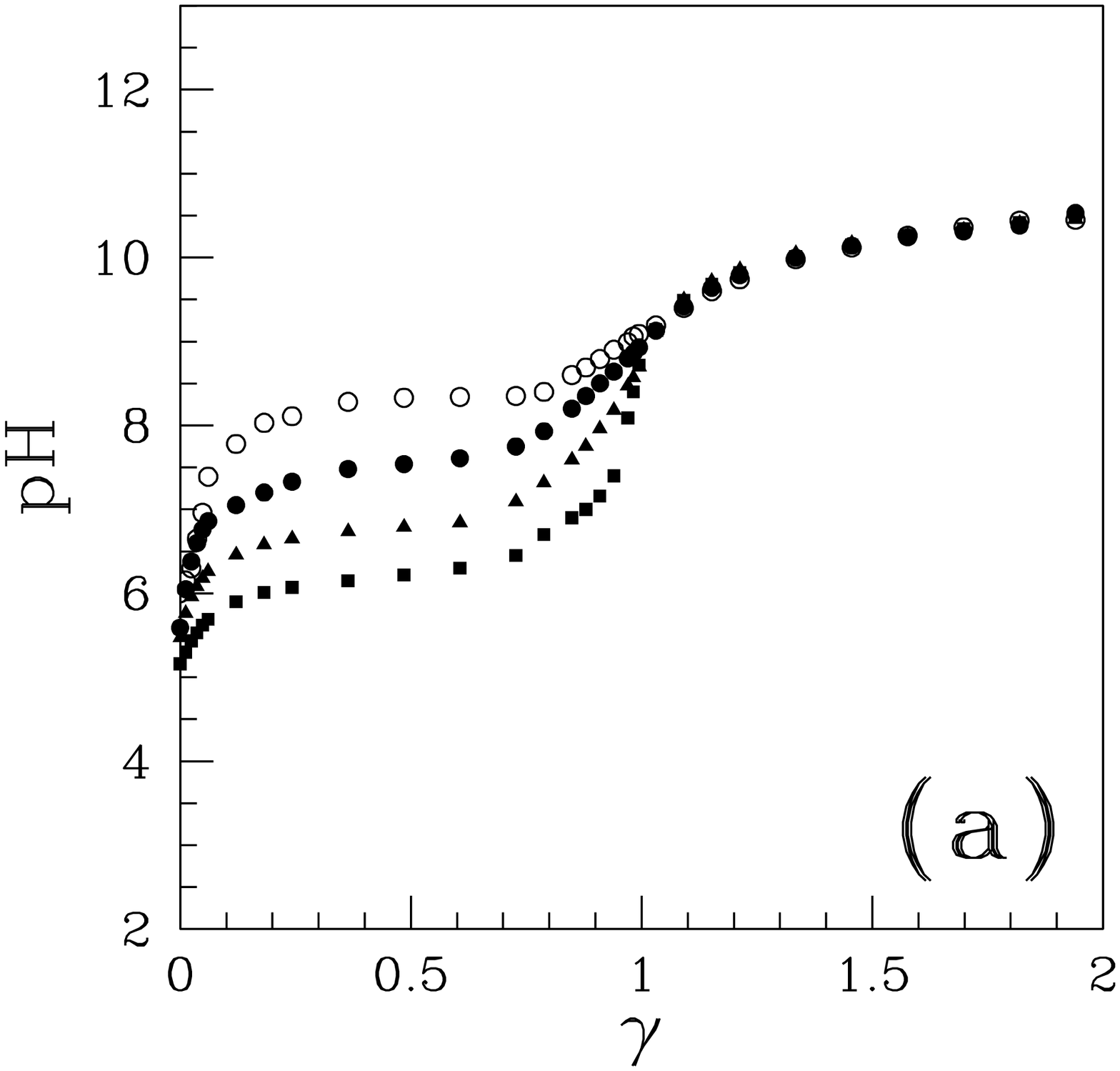}}
  \hfill
  \epsfxsize=0.4\linewidth
              \hbox{ \epsffile{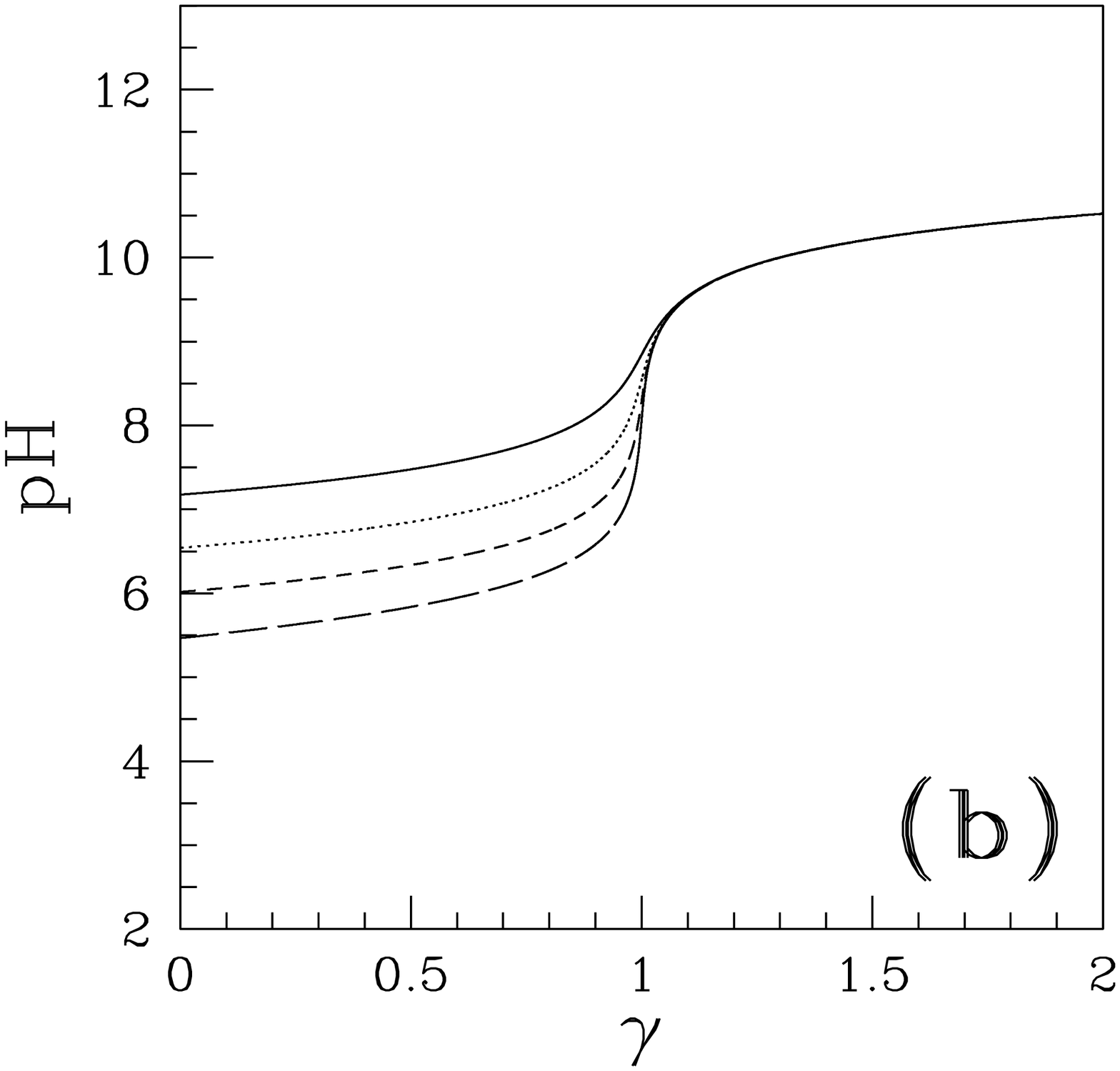}}}
\end{figure}
\vspace{1cm}
{\large Borukhov et al: Fig.~2}
\vfill

\pagebreak
\begin{figure}[tbh]
  \epsfxsize=0.4\linewidth
  \centerline{\hbox{ \epsffile{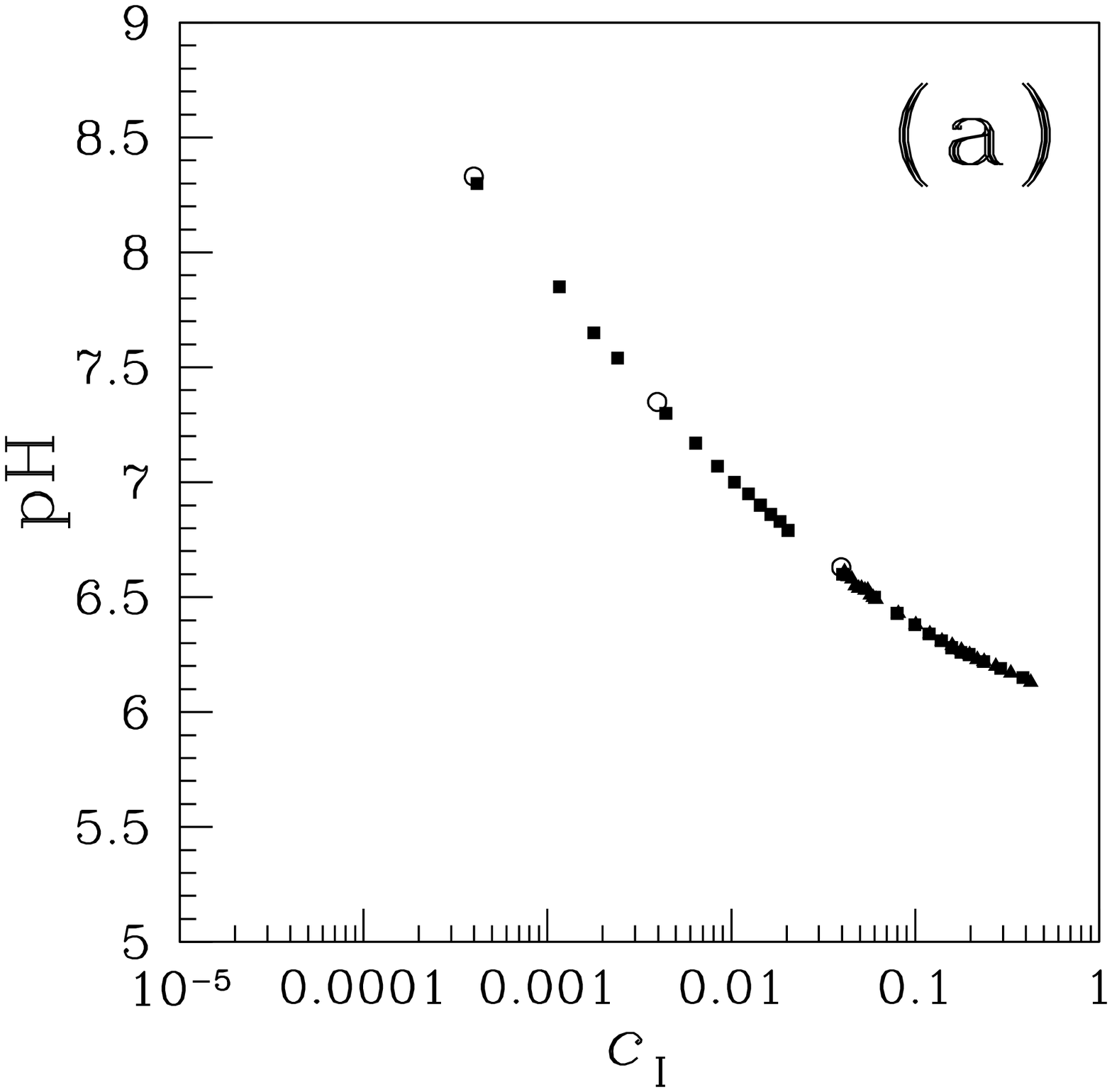}}
  \hfill
  \epsfxsize=0.4\linewidth
              \hbox{\epsffile{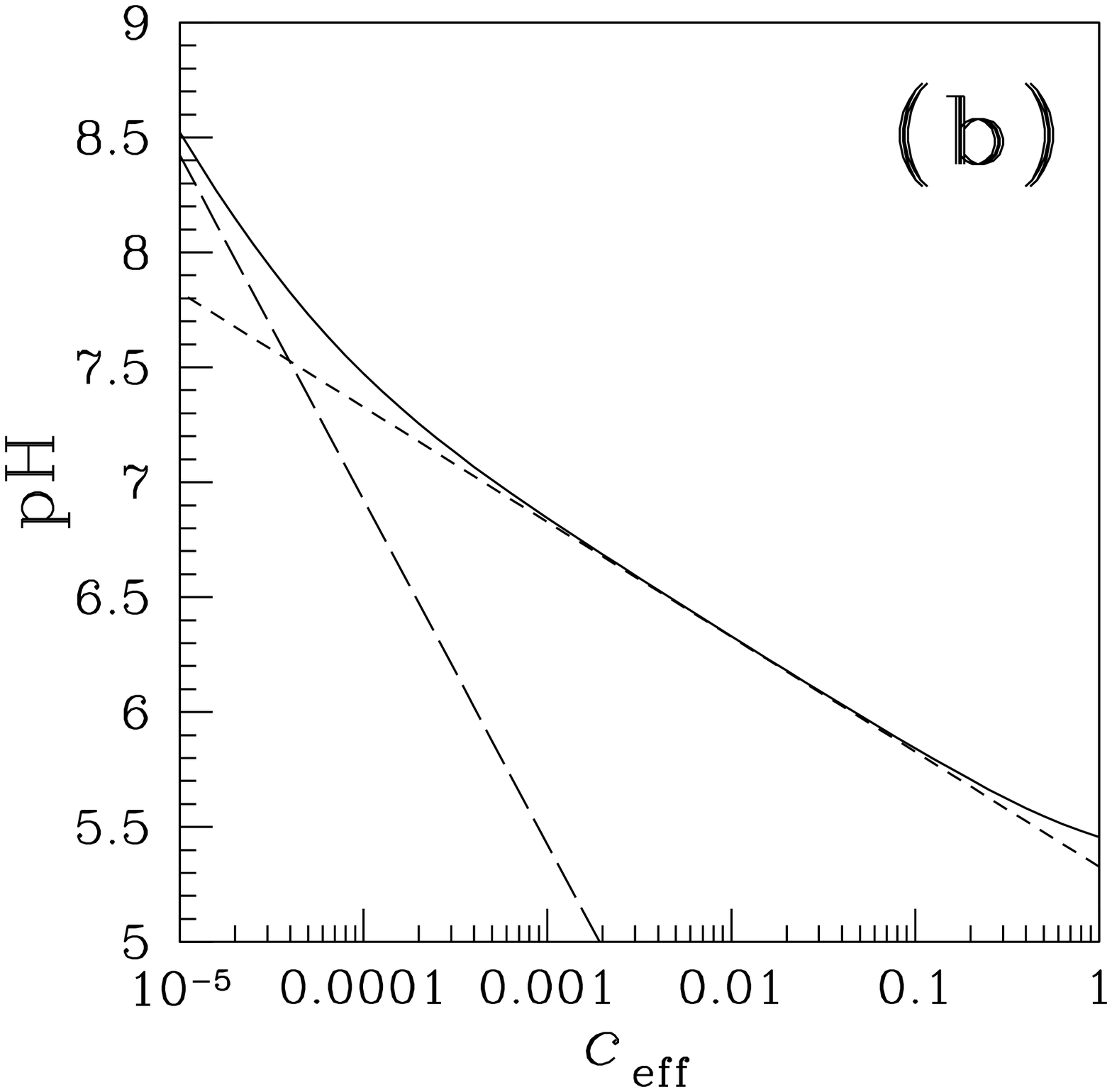}}}
\end{figure}
\vspace{1cm}
{\large Borukhov et al: Fig.~3}
\vfill

\begin{figure}[tbh]
  \epsfxsize=0.4\linewidth
  \centerline{\hbox{ \epsffile{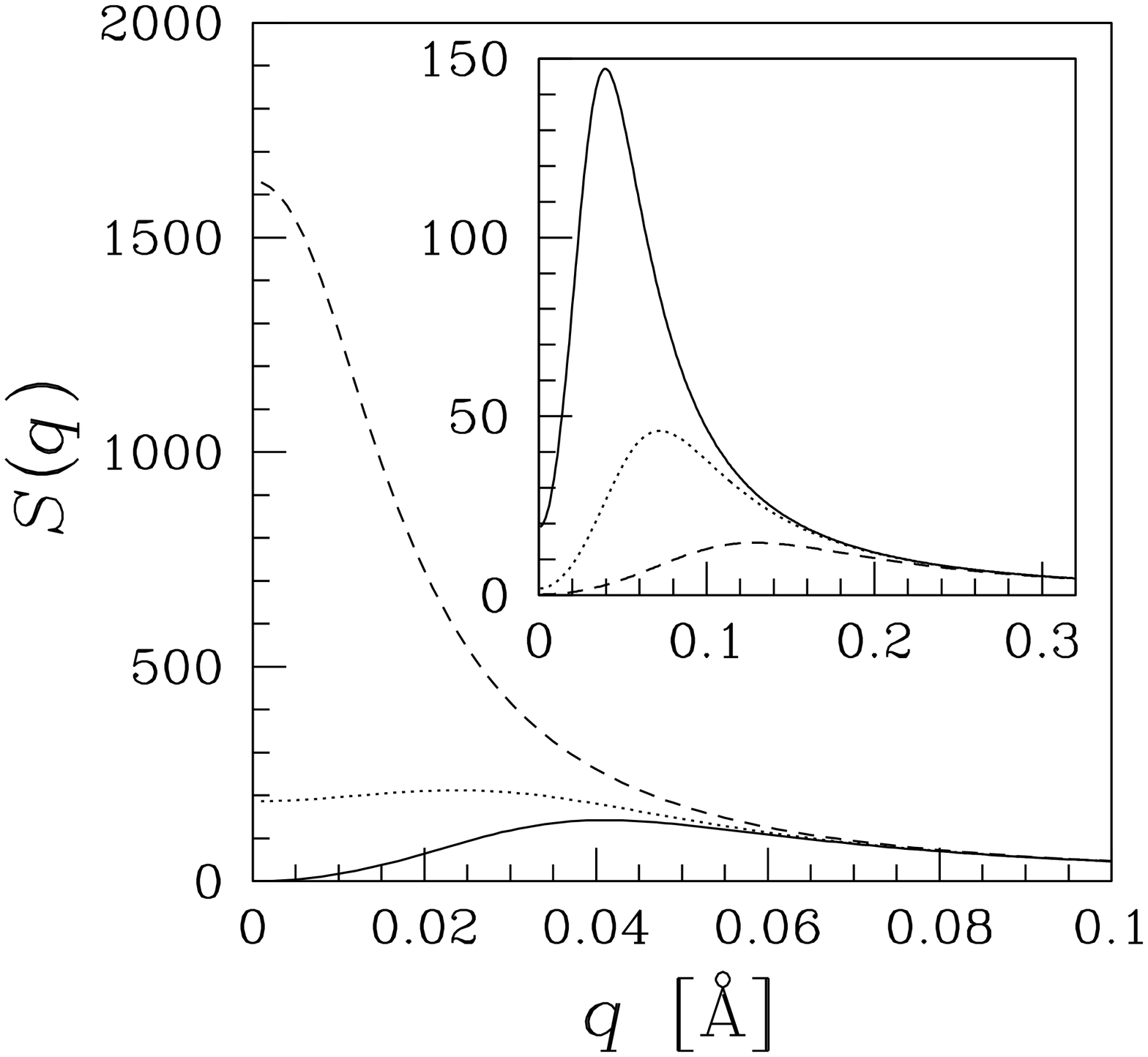}}}
\end{figure}
\vspace{1cm}
{\large Borukhov et al: Fig.~4}
\vfill


\begin{thebibliography}{99}


\bibitem{poly1}
Oosawa, F. {\em Polyelectrolytes}, Marcel Dekker: New York, 1971.

\bibitem{poly2}
De Gennes, P.G. {\em Scaling Concepts in Polymer Physics},
Cornell University: Ithaca, 1979.

\bibitem{poly3}
Barrat, J.L.;  Joanny, J.F. {\em Adv. Chem. Phys.} {\bf 1996}, 94, 1.

\bibitem{complex1}
Tanaka,T. {\em Phys. Rev. Lett.} {\bf 1978}, 40, 820.

\bibitem{complex2}
Philippova, O.E.;  Hourdet, D.;  Audebert, R.;  Khokhlov, A.R.
{\em Macromolecules} {\bf 1997}, 30, 8278.

\bibitem{complex3}
Joanny, J.F.;   Leibler, L. {\em J. Phys. (France)} {\bf 1990}, 51, 547.

\bibitem{khokhlov1}
Kokufuta, E.; Wang, B.; Yoshida, R.; Khokhlov, A.R.;  Hirata, M.
{\em Macromolecules} {\bf 1998}, 31, 6878.

\bibitem{rafael}
Raphael, E.;  Joanny, J.F. {\em Europhys. Lett.} {\bf 1990}, 13, 623;
Wittmer, J.P. {\em Ph. D thesis} (unpublished).

\bibitem{exp1}
Hasa, J.; Ilavsk\'y, M. {\em J.  Polym. Sci.} {\bf 1975}, 13, 263.

\bibitem{exp2}
Tirtaatmadja, V.; Tam, K.C.; Jenkins, R.D.  {\em Macromolecules} {\bf 1997}, 30, 3271.

\bibitem{exp3}
Tam, K.C.; Farmer, M.L.; Jenkins, R.D.; Bassett, D.R.
{\em J. Polymer Sci.: Part B: Polymer Physics} {\bf 1998}, 36, 2275.

\bibitem{th1}
Katchalsky A.; Michaeli, I. {\em J. Polym. Sci.} {\bf 1955}, XV, 69.

\bibitem{th2}
Hasa, J.; Ilavsk\'y, M.; Du\v{s}ek, K. {\em J.  Polym. Sci.} {\bf 1975}, 13, 253.

\bibitem{th3}
Brereton M.G.; Vilgis, T.A. {\em Macromolecules} {\bf 1990}, 23, 2044;
Vilgis, T.A.; Borsali, R. {\em Phys. Rev. A.} {\bf 1991}, 43, 6857;
Brereton, M.G.; Vilgis, T.A. {\em J. Phys. I (France)} {\bf 1992}, 2, 581.

\bibitem{app1}
Napper, D.H. {\em Polymer Stabilization of Colloidal Dispersions},
Academic Press: London, 1993.

\bibitem{app2}
Jenkins, R.D.; Lelong, L.M.; Bassett, D.R. in
{\it ``Hydrophilic Polymers : Performance with Environmental Acceptability''},
ed. by  Glass, J.E.,  ACS Symposium Series N° 248,
American Chemical Society {\bf 1996}, 23, 425.

\bibitem{tit1}
Mandel, M. in {\it ``Encyclopedia of Polymer Science and Engineering''}
{\bf  1988}, 11, 739.

\bibitem{tit2}
Strauss, U.P.;  Schlesinger, M.S. {\em J. Phys. Chem.} {\bf 1978}, 82, 571.

\bibitem{titration}
Atkins, P.W. {\em Physical Chemistry}, Oxford University: Oxford, 1985.

\bibitem{adv}
See, for example, {\em Adv. Polym. Sci.} {\bf 1993}, 109;  {\bf 1993}, 110.


\bibitem{epjb}
Borukhov, I.; Andelman, D.; Orland, H. {\em Eur. Phys. J. B} {\bf 1998}, 5, 869.

\bibitem{scat1}
Auvray, X.; Anthore, R.; Petitpas, C. {\em J. Phys. (France)}
{\bf 1986}, 47, 893.

\bibitem{scat2}
Heitz, C.; Raviso, M.;  Fran\c{c}ois, J. {\em Polymer}
{\bf 1999}, 40, 1637.


\end{thebibliography}
\end{document}